\documentstyle[12pt,psbox]{article}
\begin{document}

\begin{quote}

\large
\centering {\bf {Spin-Phonon Coupling in the Single-Layer Extended $t$-$J$
Model }} \\

\bigskip\smallskip
\normalsize

B. Normand, H. Kohno and H. Fukuyama

\bigskip

Department of Physics, University of Tokyo, \\
7-3-1 Hongo, Bunkyo-ku, Tokyo 113, Japan.

\end{quote}

	We consider the implications of spin-phonon coupling within the
slave-boson, mean-field treatment of the extended $t$-$J$ model of a
high-temperature superconductor. In bilayer cuprates such as YBaCuO, where
the $CuO_2$ plane is buckled, this interaction is linear in $O$ displacement
along the $c$-axis, and the coupling constant is found to be large.
The formation of a spin singlet causes additional contributions to the
phonon self-energy, and we calculate from these the superconductive phonon
anomalies. The magnitude and sign of the frequency shift and linewidth
broadening for various mode symmetries correspond well with Raman and
infra-red light scattering experiments, and with neutron scattering studies.
In the $t$-$J$ model, spin singlet formation and
superconductivity do not coincide in the low-doping regime, giving rise to
spin-gap features and a variety of temperature scales in the spin response
observed by NMR and neutron investigations. Phonon anomalies in underdoped
YBaCuO compounds indeed show evidence of spin-gap phenomena with the same
characteristic temperature, suggesting that the theory may offer the
possibility of a unified understanding of the anomalies in magnetic and
lattice properties. While the origin of the superconducting interaction is
electronic, this spin-phonon coupling affords the possibility of a small
isotope effect, and our estimate is in good agreement with recent
site-selective $O$-substitution experiments.



\newpage

\section{Introduction}

	Since the discovery of high-temperature superconductivity in
oxide-based ceramics, a wide variety of experiments has shown strong evidence
of a link between anomalies in the lattice structure and the onset of
superconductivity in these materials. A comprehensive overview of the
techniques available and their results is contained in Ref. \cite{rlehtcs}.
More specifically, a number of local probes has shown this connection:
neutron scattering studies \cite{rtejs,rarai} of the probability
distribution function for atoms in the $CuO_2$ plane show clear local
distortions at the transition temperature $T_c$ in several systems;
ion-channeling \cite{ryam} shows anomalous changes in the oscillation
amplitudes of $Cu$ and $O$ atoms not only at $T_c$, but also at
temperatures above it; EXAFS experiments \cite{roya} show changes in
the environments of both $Cu$ and apical $O$ atoms; ultrasound measurements
\cite{rnsmftk} show alterations of the elastic constants of the crystal.
Among the more general probes, there have been various reports from
Raman scattering and infrared reflectivity investigations
\cite{rliv,ralt,rham} which indicate anomalous temperature-dependences
in the frequency shift and linewidth of phonons around and above the
superconducting transition. A separate manifestation is the existence of
an isotope effect \cite{rckfmp,rfjmgs}, albeit one which appears to have
very different forms as a function of sample doping. Strikingly, infrared
reflectometry measurements of both the in-plane conductivity $\sigma_{ab}$
\cite{rrt} and the $c$-axis conductivity $\sigma_c$ \cite{rhtlbh}
also illustrate a connection between $c$-axis phonon modes
and the properties of charge carriers.

	In this work we will be particularly interested in the issue
of phonon anomalies, which are one direct consequence of the spin-phonon
coupling under consideration. Anomalies in phonon mode frequencies and
linewidths at $T_c$ in the $YBa_2Cu_3O_{7-\delta}$ (YBCO) system have been
examined since very early in the history of high-$T_c$ superconductivity
\cite{rtc}, because of the light they can shed on both the magnitude and
the symmetry (from mode polarization) of the electron-phonon interaction.
Studies by Raman spectroscopy have centered largely on the 340$cm^{-1}$
mode of planar oxygen which shows a particularly large effect in the
$O_7$ compound, and we will encounter its importance throughout the
present work. However, it is essential to note, using the well-characterized
case of this mode, that not all early measurements of the anomalies, performed
on a variety of systems, reached the same result; detailed investigations
of high-quality samples have since verified the anomaly, but have shown
that its magnitude is very sensitive to both oxygen content \cite{ralt},
falling sharply as the $O$ depletion
$\delta$ increases from zero, and to impurities \cite{rtfcc}.
Because samples with non-zero $\delta$ are non-stoichiometric, a considerable
amount of disorder may be present, acting to suppress the observed effects;
thus the stoichiometric materials $Y_2Ba_4Cu_7O_{15}$ and $YBa_2Cu_4O_8$
will represent important benchmarks of the variation in anomalies with
doping level. In other classes of the high-$T_c$ materials there remain
debates about sample purity, and rather few crystals of any single compound,
so particular care will be required in interpreting results from these.

	On the theoretical side, anomalies in the transport and magnetic
properties, observed in both normal and superconducting states of the
high-temperature superconductors \cite{rrev}, are
considered to be manifestations of a metallic state arising near the Mott
transition due to strong correlations. While there remains no consensus on
the appropriate theoretical description of the phenomena associated with
this anomalous metallic state, in this work we will pursue an
approach based on the $t$-$J$ model \cite{razr} treated by the slave-boson
mean-field theory \cite{rbza,rgjr,rkl,rSHF}. This framework has been shown
to contain ingredients essential to an understanding of both transport
properties \cite{rnlln}, including the temperature-dependence of the
resistivity and Hall coefficient, and spin excitations
\cite{rTKuF,rTKFs,rlw,rls,rmw,rul,rTKF}, such as the different
temperature-dependences
\cite{ry,rr} of the shift and rate of nuclear magnetic resonance between
high- and low-doping regions. On this last point, such models appear to
contain a consistent description of the ``spin-gap'' behavior
\cite{rr,rul,rTKF}, noted first by Yasuoka \cite{ry} and subsequently the
subject of much investigation. However, while some aspects of the link
between these features and the phonon problem have been studied theoretically,
to our knowledge there have not so far been any efforts to understand all of
the low-lying excitations on a unified basis.

	The application
of conventional electron-phonon coupling to the HTSC problem was made by
Zeyher and Zwicknagel \cite{rzz}, in a model which required a number of basic
assumptions about the nature of the system, including Fermi liquid behavior of
quasiparticles, an $s$-symmetric superconducting gap and a cylindrical Fermi
surface. These authors computed the form of the superconductive phonon
anomalies, and found that a very strong coupling constant $\lambda \simeq 3$
was required to reproduce the experimental features, which appears to be
inconsistent with later studies of transport phenomena within the same
framework \cite{rkz}. At the bandstructure level, the resonant frequencies
of many phonon modes of the YBCO structure were computed in an extensive LDA
study by Andersen {\sl {et al.}} \cite{roka}, who concluded that good
qualitative agreement with experimentally observed anomalies would be given
on application of the conventional model. These authors drew particular
attention to the ``dimpling'' modes of planar $O(2)$ and $O(3)$ atoms, not
only due to the strong anomaly in their out-of-phase oscillatory mode
(340$cm^{-1}$), but also
because they cause ``extended saddle points'' near the Fermi
surface of the electron dispersion in the region of the $(\pi,0)$ points
in $k$-space, which contribute strongly to the calculations of many physical
properties. Further studies also based on conventional electron-phonon
coupling \cite{rmz,rdvz} have given detailed descriptions of the Fano
lineshape, and pointed out how phonon anomalies may be used to deduce
the gap symmetry. We note briefly that in contrast to most of the
conventionally-based treatments \cite{rzz}, the present analysis is based
on strong correlations, encoded by the $t$-$J$ model, and takes a
simple, mean-field approach containing no further assumptions about the
nature of the system.

	We emphasize that this is not a study of a phononic mechanism for
superconductivity. In the $t$-$J$ and related models, quasiparticle pairing
is the consequence of the spin interaction, and so its origin is entirely
electronic. In this work we investigate the effects of phonon oscillations
of the host lattice on the parameters of the model itself, showing that
they cause an effective spin-phonon coupling. As explained in the following
section, while the coupling constant is large, the net interaction depends
also on the magnitude of the phonon oscillation, and is found
to be small. However, its results are clearly observable in a variety of
experimental properties, which may in turn shed much light on the behavior
of the system itself.

	The structure of this paper is as follows. In section 2 we
introduce the concept of a coupling between the spin and lattice degrees
of freedom, and deduce the quantitative form of the interaction vertices.
In section 3 is given a complete account of the application of the
single-layer theory to phonon anomalies in $YBa_2Cu_3O_7$. In section
4 we present a qualitative discussion of the extension of the same ideas
to underdoped YBCO materials, leading to a connection with the mean-field
phase diagram and spin-gap behavior. Section 5 contains an analysis of
the isotope effect, and section 6 a concluding discussion. Some of the
key results of this paper have been presented in Ref. \cite{us}.

\section{Spin-Phonon Coupling}

	We consider the spin-phonon coupling arising naturally within the
extended $t$-$J$ model of a single $CuO_2$ layer, which has been shown
\cite{rTKFs,rTKF} at the mean-field level to give a good account of many
features of the spin excitations in both $La_{2-x}Sr_xCuO_4$ (LSCO) and
$YBa_2Cu_3O_{7-\delta}$ (YBCO), based on realistic Fermi surface shapes
for each class of material. The Hamiltonian is
\begin{equation}
H = - \sum_{ij, \sigma} t_{ij} a_{i \sigma}^{\dag} a_{j \sigma} +
\sum_{ \langle ij \rangle } J_{ij} {\bf S}_i {\bf {.S}}_j ,
\label{eh}
\end{equation}
where the Hilbert space is that without double occupancy, $t_{ij}$ corresponds
to the transfer integrals used to reproduce the Fermi surface, and the
superexchange interaction $J_{ij}$, which is usually taken to be a constant
at fixed doping, is assumed to be finite only between nearest neighbors.

	In YBCO, the $Cu O_2$ layer is ``buckled'', by which is meant
that the oxygen atoms $O(2)$ and $O(3)$ lie out of the plane of the $Cu$
atoms, as shown schematically in Fig.~1. Let $u_0 + u_{i}^{\alpha}$ represent
the magnitude of the oxygen displacement along the $c$-axis, with $u_0$ the
equilibrium buckling and $u_{i}^{\alpha}$ the phonon coordinate, in which
$i$ refers to the $Cu$ sites on the square lattice,
and $\alpha$ to either $O(2)$ ($\alpha = x$) or $O(3)$ ($\alpha = y$)).
Then $t_{ij}$ and $J_{ij}$ between nearest-neighbor $Cu$ sites have
contributions linear in $u_{i}^{\alpha}$ which may be written as
\begin{eqnarray}
t_{i,i+\alpha} & = & t [ 1 - \lambda_t (u_{i}^{\alpha}/a) ], \label{elcce1} \\
J_{i,i+\alpha} & = & J [ 1 - \lambda_J (u_{i}^{\alpha}/a) ], \label{elcce2}
\end{eqnarray}
with $a$ the distance between $Cu$ sites, and $\lambda_t$ and $\lambda_J$
the corresponding coupling constants. Microscopic estimates of $\lambda_t$
and $\lambda_J$ are somewhat involved, and we proceed to detail the steps
required in their derivation.

	The degree of buckling may be deduced from the structural data of
Ref. \cite{rgui}, which shows the equilibrium $O$ displacement to be
$u_0 = 0.256 {\rm {\AA}}$; this data was taken at 10K, but the
temperature-dependence is thought to be very weak, even at the
superconducting transition. We neglect the 5\% anisotropy between $a$-
and $b$-axes in the pseudo-tetragonal system, and thus continue also
with the assumption that the atomic levels of the $\sigma$- and $\pi$-orbitals
on $O(2)$ and $O(3)$ are the same. Although the degree of buckling is small,
$u_0/a \ll 1$, its inclusion is crucial in providing a coupling which is
strong and linear.

	The dependence on interatomic separation of the transfer integral
$t_{\sigma}$ ($t_{\pi}$) between $Cu$:$d_{x^2 - y^2}$ and $O$:$p_{\sigma}$
($O$:$p_{\pi}$) is given, following Ref. \cite{rHarrison}, by
\begin{eqnarray}
t_{\sigma} & = & \frac{\sqrt{3}}{2} V_{pd \sigma} \left[ 1 -
\frac{3}{2} \left( \frac{2 u}{a} \right)^2 \right] + \left( \frac{2 u}{a}
\right) V_{pd \pi} \label{etsp} \nonumber \\ & = & \frac{\sqrt{3}}{2}
V_{pd \sigma} \left[ 1 - 2.03 \left( \frac{2 u}{a} \right)^2 \right]
\nonumber \\ & = & \frac{\sqrt{3}}{2}
V_{pd \sigma}^0 \left[ 1 - 3.78 \left( \frac{2 u}{a} \right)^2 \right] \\
t_{\pi} & = & \frac{\sqrt{3}}{2} \frac{2 u}{a} V_{pd \sigma}^0 \left(
1 - \frac{V_{pd \pi}^0}{V_{pd \sigma}^0} \right) \;\; = \;\; 1.53 \left(
\frac{2 u}{a} \right) .
\end{eqnarray}
Here $\mbox{$\frac{\sqrt{3}}{2}$} V_{pd \sigma}$ ($V_{pd \pi}$) is the
transfer integral between $d_{x^2 - y^2}$ ($d_{xz}$) and $p_{\sigma}$
($p_{z}$) orbitals with separation $d = \left( (\mbox{$\frac{1}{2}$} a)^2 +
u^2 \right)^{1/2}$, directed along ${\hat x}$ according to definition, and
we have taken the distance-dependence to be
\begin{equation}
V_{pd} (d) = V_{pd}^0 \left( \frac{d}{d_0} \right)^{- 7/2} .
\label{ehbdd}
\end{equation}
By writing $u = u_0 + \delta u$, where $\delta u$ ($\equiv u_{i}^{\alpha}$)
denotes the oscillation
amplitude, for the value of $u_0$ above one obtains to lowest order
$t_{\sigma} (u) / t_{\sigma} (u_0) = 1 - 2.03 \delta u / a$ and
$\left( t_{\pi} (u) - t_{\pi} (u_0) \right) /$ $ t_{\sigma} (u_0) =
3.06 \delta u / a$.

	The final requirement is the dependence on $t_{\sigma}$ and
$t_{\pi}$ of $t_{ij}$ and $J_{ij}$, and here we must consider in detail a
model of the $CuO_2$ plane. Following Eskes and Jefferson \cite{rej},
$J_{ij}$ can be given by the perturbative expression
\begin{equation}
J = \frac{4 \left( t_{pd,{\sigma}}^4 - 2 t_{pd,{\sigma}}^2 t_{pd,{\pi}}^2
\right)}{( \Delta_{CT} + U_{pd} )^2} \left[ \frac{1}{U_d} + \frac{2}{2
\Delta_{CT} + U_p} \right] ,
\label{ej4o}
\end{equation}
where $U_d$ and $U_p$ are on-site interactions for the participating $Cu$
and $O$ orbitals, $U_{pd}$ is a Coulomb interaction between holes on
neighboring sites and $\Delta_{CT}$ is the charge transfer energy.
Application of this formula alone yields a very large coupling constant
$\lambda_J = 10.4$, because of the quartic power law and because the
contributions from $\sigma$- and $\pi$-hopping processes in the numerator
combine ferromagnetically. However, use of the lowest-order perturbation form
(\ref{ej4o}) is not well justified, as there are no small parameter ratios,
and in particular the direct $O$-$O$ hopping integral $t_{pp}$ will have
a strong effect. A detailed investigation of the influence of higher-order
terms \cite{rej} has found for the parameters of the $Cu O_2$ layer the
effective relationship $J \propto t_{pd}^x$, with $x \simeq 2.3$, when
all overlap integrals are taken to vary uniformly ($t_{pp} \propto
t_{pd}^{2/3}$ \cite{rHarrison}) and $\Delta_{CT}$ is held fixed. However,
this result depends upon the
variation of the transfer integrals with the bond distortion associated
with each phonon mode, and for a mode in which $t_{pp}$ remains constant
($A_{1g}$ and $A_{2u}$ symmetries, section 3), the effective power is
reduced to $x \simeq 1.7$ \cite{repc}.

	The case of the hopping terms $t_{ij}$ is complicated further by
the fact that it requires consideration of the many inter-site transfer
integrals which contribute to the hopping of a Zhang-Rice singlet, and not
simply of the motion of a hole. This problem has been considered within a
cell-perturbation method \cite{rjef}, and the results for the effective
power in $t \propto t_{pd}^y$ are $y \simeq 1.0$ for uniform variation,
and $y \simeq 0.7$ when $t_{pp}$ is held constant \cite{repc}. In this
study, our primary aim is to elucidate the effects of spin-phonon coupling
in the $t$-$J$ model at a semi-quantitative level, with attention paid to
the vertex magnitudes mainly to ensure that the consequences discussed are
not irrelevant to experiment. Thus for the current purposes
we accept possible errors of order 20-30\% in adopting the
values $x = 2.0$ and $y = 1.0$, so that $\lambda_J = 5.2$ and $\lambda_t
= 2.6 = \mbox{$\frac{1}{2}$} \lambda_J$, and errors of similar magnitude
incurred by neglecting any modulation of the extended singlet hopping
terms. [In fact, the first set are likely to lead to an overestimate, while
the second, in the most naive approximation, would give a similar
underestimate, although this depends on the mode symmetry.] An independent
indication of the validity of these values is provided by the agreement
they yield with the measured isotope shift in YBCO \cite{rzea} (section 5).

	In this discussion we have disregarded changes in the charge-transfer
energy ${\Delta_{CT}}$ as the bond length varies, because we believe this
to be appropriate for the deformation processes associated with a phonon
oscillation, which are both local and screened. This situation is to be
contrasted with that in experiments in which the bond length is made to vary
by application of hydrostatic pressure \cite{rssws}, or by atomic substitution
\cite{rTokura,rotm}, where one finds the weaker relation $J \propto
d^{- \alpha}$, $4 < \alpha < 6$. In these cases long-range forces are
brought into play, the Madelung energy of the system is altered, and it is
found that a $d$-dependence of $\Delta_{CT}$ is required to account for
experiment \cite{rsa}.

	In the slave-boson treatment of the $t$-$J$ model (\ref{eh}),
the spin and charge degrees of freedom carried by the quasiparticles are
represented by the explicit decomposition $a_{i \sigma} = f_{i \sigma}
b_{i}^{\dag}$, where $f_{i \sigma}$ is a fermionic spinon, in terms of
which the spin is given by ${\bf S}_i = \mbox{$\frac{1}{2}$}
f_{i \alpha}^{\dag} {\vec {\sigma}}_{\alpha \beta} f_{i \beta}$, and $b_i$
is a bosonic hole, or holon. With these operators, the coupling
in $H$ (\ref{eh}) of the phonon coordinate $u_{i}^{\alpha}$ to the
spin degrees of freedom may be expressed in the mean-field approximation as
\begin{eqnarray}
& & - \sum_{i} \sum_{\alpha = x,y} \left\{ t \lambda_t \mbox{$\left(
\frac{u_i^{\alpha}}{a} \right)$} \langle b_i b_{i + \alpha}^{\dag} \rangle
\chi_{i,i + \alpha} \right. \label{ejss} \\ & &
\left. + \mbox{$\frac{3}{8}$} J \lambda_J \mbox{$\left(
\frac{u_i^{\alpha}}{a} \right)$} \left[ \langle \chi_{i,i + \alpha}^{\dag}
\rangle \chi_{i,i + \alpha} + 2 \langle \Delta_{ij}^{\dag} \rangle
\Delta_{ij} + h.c. \right] \right\} \nonumber ,
\end{eqnarray}
where $\chi_{i,j} = \sum_s f_{i s}^{\dag} f_{j s}$ and $\Delta_{ij} = \left(
f_{i \uparrow} f_{j \downarrow} - f_{i \downarrow} f_{j \uparrow} \right)
/ \sqrt{2} $. The first two terms, containing $\chi_{i,j}$, will contribute
in the normal, or uniform RVB \cite{rTKF}, state, and will be termed the
$u$-RVB vertex, while the last appears only below the singlet RVB
transition, so constitutes an $s$-RVB vertex.
At the onset of singlet order there will be additional
contributions to any physical quantity due to the appearance of a finite
particle-particle vertex represented by the 3rd term, as well as to changes
in the quasiparticle propagators joining the particle-hole terms. In this
analysis of the $t$-$J$ model \cite{rTKF} we do not include a density-density
($n_i n_j$) decoupling of the spin interaction term
$J {\bf S}_i {\bf {.S}}_j$, because it has been found \cite{rmpc}
to have a small coefficient in the derivation from the $d$-$p$ model.
In principle the phonon-holon coupling vertex, which is also contained in
the $t$ term, may contribute to physical processes, but this will be
neglected because in the normal state a boson polarization term vanishes
at $q = 0$, while in the Bose-condensed state corresponding to true
superconductivity in this model the holon has no dynamics.

	While the coupling constants $\lambda_t$ and $\lambda_J$ have been
found to be large, the overall strength of the spin-phonon coupling depends
also on the magnitude of the phonon oscillations $u_{i}^{\alpha} / a$, and
so remains small. It can be estimated from the root mean square fluctuations
of the relative magnitudes of $t$ and $J$ ((\ref{elcce1}) and
(\ref{elcce2})), by using $\sqrt{ \langle ( t_{i,i+\alpha} - t )^2
\rangle } / t = \frac{\lambda_t}{a} \sqrt{ \langle (u_{i}^{\alpha})^2
\rangle }$, and similarly for $J$. The r.m.s. fluctuation of $\delta u
\equiv u_{i}^{\alpha}$ is given by $\langle
\delta u^2 \rangle = \mbox{$\frac{\hbar}{2 M \omega_0}$}$, where $\omega_0$
is taken for the phonon frequency of interest and $M$ is the mass of the
$O$ atom. Taking as a typical frequency that of the 340$cm^{-1}$ mode, to
which we will devote most attention in the following section, $\langle \delta
u^2 \rangle = ( 0.055 {\rm {\AA}} )^2$, and the relative fluctuations in $t$
and $J$ are
3.8\% and 7.5\% respectively. These values appear surprisingly large, but
are consistent with the estimates of Haas {\sl {et al.}} \cite{rhdrmn},
who considered only the effect of a fluctuating $J$ term on the linewidth
in electronic Raman scattering. In what follows we will find also that the
changes in physical quantities arising from the combined effects of these
fluctuations are of the order of their square.

\section{Phonon Anomalies}

	Having derived a coupling between spin and lattice modes, we
consider first its modifcation of phonon dynamics. In Fig.~2 are shown
the four mode symmetries of a $CuO_2$ bilayer in which the $O(2)$ and $O(3)$
atoms may oscillate in the $c$-direction. We will use throughout the
nomenclature of the $D_{4h}$ symmetry obeyed by the pseudo-tetragonal
$CuO_2$ planar unit in the slightly anisotropic YBCO system, and refer
the reader to Ref. \cite{rtc} for full details. Here we will be concerned
mainly with the 340$cm^{-1}$ $B_{1g}$ mode
in $YBa_2Cu_3O_7$, which is an out-of-phase oscillatory motion of
the planar oxygen atoms ($u_{i}^x = - u_{i}^y$), and is also out of phase
between the planes of the bilayer \cite{rtc}. This mode has attracted
experimental interest because it shows the largest effects in the
superconducting state, and there are available both detailed Raman
\cite{rliv,ralt,rham} data, which is taken at wavevector $q = 0$, and
inelastic neutron scattering results \cite{rprperh}, which can probe all
$q$. The primary quantity of experimental interest is the $T$-dependence
of the frequency shift and linewidth at $q = 0$, in addition to
which we will illustrate the dependence of the results on mode frequency
and wavevector. The 193$cm^{-1}$ $B_{2u}$ mode is identical at the
single-layer level, although in the real crystal the atomic motions are
in-phase between layers, and so is also amenable to description by the
same model at this lower frequency. Because this mode is infrared-silent,
it may be studied only by neutron scattering, and it has recently been
investigated in $Y Ba_2 Cu_3 O_7$ \cite{rhs}. We will also show the results
of the single-layer model for the in-phase ($u_{i}^x = u_{i}^y$) $A_{1g}$-
and $A_{2u}$-symmetric oscillations, and discuss the relevance of the
model to experiment for these.

\subsection{Theory}

	The effect of the coupling on the dynamical properties of the
phonon is calculated from the lowest-order spinon polarization correction
to the phonon self-energy $\Pi_{sp} ({\bf q}, i \omega_n)$, shown in Fig.~3;
the full phonon propagator is given by $\Pi_{ph}^{-1} ({\bf q}, i \omega_n)
= \Pi_{0}^{-1} ({\bf q}, i \omega_n) - \Pi_{sp}^{-1} ({\bf q}, i \omega_n)$,
where
the bare propagator has the conventional form $\Pi_{0} ({\bf q}, i \omega_n)
= 2 \omega_0 / i \omega_{n}^2 - \omega_{0}^2$. Thus when $|${\sl Re}$\,
\Pi_{sp}| \ll \omega_0$, the superconductivity-induced phonon frequency shift
is given to lowest order by $\delta \omega = ${\sl Re}$\, \Pi_{sp}$, and the
correction to the linewidth $\Gamma$ is $\delta \Gamma = ${\sl Im}$ \,
\Pi_{sp}$. Second-order perturbation theory in terms of the interaction
strength results in a frequency shift which is given
at $q = 0$ by
\begin{equation}
\delta \omega = c \left( \lambda_J J \right)^2 \frac{4}{N} \sum_{k} F_k
\frac{1}{\omega^2 - \left( 2 E_k \right)^2} \frac{\tanh \left( \frac{E_k}{2 T}
\right)}{E_k} ,
\label{edo}
\end{equation}
where $c = \mbox{$\left( \frac{3}{4 a} \right)^2$} \langle \delta u^2
\rangle$, with $\langle \delta u^2 \rangle$ defined at the end of section 2,
and has the value $c = 1.18 \times 10^{-4}$ for the $B_{1g}$
phonon. $E_k$ is defined by $E_k = \left[ \xi_{k}^2 + \Delta_{k}^2
\right]^{1/2}$, with $\xi_k$ the spinon band energy relative to the
chemical potential, and $\Delta_k = - \mbox{$\frac{3 \sqrt{2}}{4}$}
J \Delta (\cos k_x - \cos k_y)$ the singlet order parameter, which has
been shown to have $d_{x^2 - y^2}$-wave symmetry in the lowest-energy
state in the present framework \cite{rgjr,rkl}; the negative branch given
by $- E_k$ is illustrated in Fig.~4 for the parameters of the YBCO system
used in this study. Finally, $F_k$ is a form factor which depends on the
mode symmetry,
\begin{eqnarray}
F_k & = & 2 \Delta^2 \left( \gamma_k \xi_{k} + \mbox{$\frac{3 J}{4}$}
{\bar {\chi}} \eta_{k}^2 \right)^2 \;\;\;\;\;\;\;\;\;\; B_{1g}, \, B_{2u}
\; {\rm {modes}} \label{eff} \nonumber \\ F_k & = &
2 \Delta^2 \eta_{k}^2 \left( \xi_{k} + \mbox{$\frac{3 J}{4}$}
{\bar {\chi}} \gamma_k \right)^2
\;\;\;\;\;\;\;\;\;\; A_{1g}, \, A_{2u} \; {\rm {modes}} ,
\end{eqnarray}
where $\gamma_k = \cos k_x + \cos k_y$, $\eta_k = \cos k_x - \cos k_y$,
$\Delta = \langle \Delta_{ij} \rangle$ and ${\bar {\chi}} \equiv
\langle \chi_{ij} \rangle + \frac{2 t \delta}{3 J}$ is written to contain
the effects of the particle-hole vertices from both $J$ and $t$ terms;
$\delta$ is the hole concentration.
In Eq. (\ref{eff}) one has $F_k \propto \Delta^2$, so that there is no
phonon energy correction due to spin coupling in the normal state: this
result applies only at $q = 0$ for the optic phonon modes under
consideration. By a similar analysis, the linewidth broadening is given by
\begin{equation}
\delta \Gamma = - c \left( \lambda_J J \right)^2 \frac{\pi}{N} \sum_{k} F_k
\left[ \delta ( \omega - 2 E_k ) - \delta ( \omega + 2 E_k ) \right]
\frac{\tanh \left( \frac{E_k}{2 T} \right)}{E_{k}^2} .
\label{edg}
\end{equation}

\subsection{Results}

	For the following calculations we use the self-consistent solutions
to the mean-field equations of the extended $t$-$J$ model \cite{rTKFs} for
the temperature-dependence of the parameters (${\bar {\chi}}$, $\Delta$,
$\mu$) for

\smallskip
\noindent
i) the transfer integrals
appropriate to YBCO: $t = 4 J$, $t^{\prime} = - \mbox{$\frac{1}{6}$} t$ and
$t^{\prime \prime} = \mbox{$\frac{1}{5}$} t$ (corresponding to hopping
processes between nearest, next-nearest and third-neighbors, respectively),
giving the spinon dispersion shown in Fig.~4, and

\smallskip
\noindent
ii) a doping level $\delta = 0.2$, corresponding approximately to optimal
doping, or $Y Ba_2 Cu_3 O_7$ in this model \cite{rTKF}.

\smallskip
\noindent
We note here that in the $t$-$J$ models with the superexchange $J$ calculated
in the random-phase approximation (RPA), the doping cannot be taken to be a
parameter which may be
varied to reproduce experimental observations. The chosen doping level is
taken neither too near nor too far from the antiferromagnetic instability
of the system, so that the quantities calculated may be considered to be a
reasonable reflection of the effects of spin fluctuation enhancement, without
being singular \cite{rTKF}. Thus we will not have any semi-quantitative
statements to make with regard to doping-dependence.
Throughout the calculations we have assumed a Lorentzian broadening
of the spinon spectrum $\gamma = 0.12 k_B T_{RVB}$, and we find that the
results depend very little on value of this parameter: specifically, a
reduction of $\gamma$ by a factor of 10 is required to enhance the
frequency shift by a factor of 2, for mode frequencies close to the
maximum shift (Fig.~5(a) below).

\subsubsection*{$\omega$-Dependence}

	Using the mean-field parameters, the frequency-dependence of
$\delta \omega$, given by Eq. (\ref{edo}), has been evaluated numerically
at $T/T_{RVB} = 0.2$ and $T/T_{RVB} = 0.8$, where $T_{RVB} = 0.069 J$ is
the onset temperature for the singlet RVB order parameter $\Delta$, and
the results are shown by the $\circ$ symbols in Fig.~5 for $q = 0$.
The change in sign with frequency for $\delta \omega$ of the $B_{1g}$ mode
is a result of the $\omega$-dependent denominator in Eq. (\ref{edo}). The
frequency of the crossing is an approximate measure of the value of $2
\Delta_k (T)$ near the $(\pi,0)$ points, where the gap is maximal, as these
regions are favored by the $B_{1g}$ symmetry. As the temperature is
increased towards $T_{RVB}$, the $\omega$-dependence of $\delta \omega$
remains qualitatively the same, but the sign-change occurs at lower
frequencies as the gap magnitude falls. The experimental mode frequency
$\omega_{ph} = 340cm^{-1} \simeq 0.3 J$ for the $B_{1g}$ mode is near, but
just below, the frequency of the crossing at low temperatures, so this
mode can be expected to show a maximal effect.

	The linewidth broadening $\delta \Gamma$ is shown in Fig.~6 for the
same two temperatures, and has the form to be expected for an imaginary part
of the quantities whose real part is shown in Fig.~5, namely a peak at the
frequency of the sign-change in $\delta \omega$. Note that because the gap
is $d$-symmetric (point nodes on the Fermi surface) there is no region at
low frequencies  where the imaginary part vanishes.

\subsubsection*{$T$-Dependence}

	The temperature-dependence of
$\delta \omega$ is shown in Fig.~7 for $q = 0$ and the frequency choices
$\omega_0 / J = 0.15$ and $0.25$, which are indicated by arrows in
Fig.~5: the former is in the regime where the mode frequency is considerably
less than $2 {\rm {Max}} [ \Delta_k (T = 0) ] \equiv $ ``$2 \Delta$'',
and shows a sharp transition; the latter value has reasonable quantitative
agreement with the $340cm^{-1}$ mode, and shows clearly that as the mode
frequency approaches $2 \Delta$, the shift in frequency occurs at a
temperature somewhat below $T_{RVB}$. This corresponds well to the
observations of Ref. \cite{ralt}, where the full frequency shift develops
over a range of temperatures below the onset.
In fact, at the value $\omega_0 = 0.3J$, closest to the exact experimental
mode frequency $\omega_{ph}$, there is a small, positive frequency shift
because $\omega_0 > 2 {\rm {Max}} [ \Delta_k (T = 0) ]$; however, we
illustrate the anomalous effects for a mode frequency just below this value,
as is shown to be the case in experiment, and in general regard the
degree of correspondence between model and experimental mode and gap energies
as a success of the initial mean-field formulation.

	Comparison with the experimental result \cite{ralt} for the
temperature-dependent broadening of the $B_{1g}$ mode shows extremely
good agreement in functional form and in magnitude ($\delta \omega \simeq
\delta \Gamma \simeq 0.01 \omega_0$) for a frequency $\omega_0$ close to
$\omega_{ph}$ (Fig.~8(b)). At lower frequencies, away from the ``resonance''
energy set by the low-temperature gap, $\delta \Gamma$ is suppressed
(Fig.~8(a)). Note that in Figs.~7(a) and ~8(a)
(and~10(a)) we retain the labelling $B_{1g}$ for the symmetry of the phonon
distortion, although in fact the frequency is chosen to be close to that
of the $B_{2u}$ mode.

	Quantitatively, the magnitude of the effects given by the model
with the chosen values of $\lambda_t$ and $\lambda_J$ is within a factor
of 1.5 of the Raman measurements \cite{ralt} on $Y Ba_2 Cu_3 O_7$.
In the light of the approximations detailed in the preceding section,
such correspondence may be taken as justification for these, and appears
eminently satisfactory within a mean-field treatment using no adjustable
parameters. We note also that the magnitude of the anomalies in frequency
and broadening appear to drop quickly with doping $\delta$, and that
neutron measurements at $q = 0$ \cite{rprperh} suggest a somewhat smaller
frequency shift than is seen by Raman scattering.

\subsubsection*{Gap Symmetry}

	The frequency shift and linewidth results obtained for the $B_{1g}$
and $B_{2u}$ modes
are consistent only with predominantly $d_{x^2 - y^2}$-wave singlet
pairing, as this has a constructive combination with the phonon symmetry
in the expression (\ref{eff}) for the form factor. By contrast, a
conventional extended $s$-symmetric gap is found to give an immeasurably
small frequency shift for both $B$- and $A$-type mode symmetries, because
the form factors ($F_k$) are small for $B$-symmetry, whil both thermal and
the form factors are small for $A$. While a
pairing state characterized by an intermediate phase $\Delta_y / \Delta_x
= {\rm e}^{i \phi}$ \cite{rSHF,rnkf}, with $\phi$ different from but close
to $\pi$, so that the system is close to $d$-symmetry, cannot be excluded by
the present analysis alone, it appears that all forms of anisotropic
$s$-wave gaps are ruled out by their antiphase combination with the phonon
symmetry (see also Ref. \cite{rdvz}).

\subsubsection*{${\bf q}$-dependence}

	We have extended our analysis to the case where the phonon has a
finite wavevector, and will consider here only the variation of the frequency
shift, $\delta \omega$, as this has been measured. $\delta \omega ({\bf q},
\omega, T)$
exhibits a wealth of features as ${\bf q}$ is varied, because the quantitative
effects are sensitive to the detailed shape of the spinon dispersion $E_k$
(Fig.~4), as well as to the bare phonon frequency as illustrated above.
When the phonon has finite wavevector ${\bf q}$, one may show that the
components of ${\bf u} ({\bf q})$ obey the relations $u_x ({\bf q}) = \pm
u_y ({\bf q})$ for every ${\bf q}$ as a consequence of $u_{i}^{x} = \pm
u_{i}^{y}$, and the expression (\ref{edo}) takes the more general form
\begin{eqnarray}
\delta \omega ( {\bf q} ) & = & - c \left( \lambda J \right)^2 \frac{2}{N}
\sum_{k} \left\{
F^{(+)} \left( f ( E_{+} ) - f ( E_{-} ) \right) \frac{ \left( E_{+}
- E_{-} \right) }{\omega^2 - \left( E_{+} - E_{-} \right)^2} \right.
\label{eqdo} \nonumber \\ & & \;\;\;\;\;\;\;\;\;\;\;\;\;\;\;\;\;\;\;\;\;\;
\;\;\;\; - \left. F^{(-)} \left( 1 - f ( E_{+} ) - f ( E_{-} ) \right) \frac{
\left( E_{+} + E_{-} \right) }{\omega^2 - \left( E_{+} + E_{-} \right)^2}
\right\},
\end{eqnarray}
where the form factors for the normal and anomalous processes are given by
\begin{eqnarray}
F^{(\pm)} & = & \frac{1}{2} \left( c_{1}^2 + c_{3}^2 \right) \; \pm \;
\frac{1}{2 E_{+} E_{-} } \left[ (c_1 \Delta_{+} + c_3 \xi_{+} )
(c_1 \Delta_{-} + c_3 \xi_{-} ) \right. \label{eqff} \nonumber \\ & &
\;\;\;\;\;\;\;\;\;\;\;\;\;\;\;\;\;\;\;\;\;\;\;\;\;\;\;\;\;\;\;\;\;\;\;\;
- \left. ( c_1 \xi_{+} - c_3 \Delta_{+} )
( c_1 \xi_{-} - c_{3} \Delta_{-} ) \right],
\end{eqnarray}
in which the subscripts $\pm$ denote momentum labels ${\bf k} \pm
\mbox{$\frac{1}{2}$} {\bf q}$ throughout, and $( c_1, c_3 ) = ( \sqrt{2}
\Delta_0 \gamma_k , \bar{\chi} \eta_k)$ for the $B_{1g}$ mode, and
$( \sqrt{2} \Delta_0 \eta_k , \bar{\chi} \gamma_k)$ for the $A_{2u}$ mode.
We assume for simplicity that the phonon is non-dispersive, {\sl {i.e.}}
that the mode frequency changes little with $q$, and this appears to be
borne out by experiment.

	In Fig.~9 are shown the frequency shifts at ${\bf q} = (0.2,0) \pi$
as a function of $\omega$ at low $T$ ((a), {\it cf.} Fig.~5(a)), and of $T$ at
$\omega_0 = 0.25 J$ ((b), {\it cf.} Fig.~7(b)). In the $\omega$-dependence
there is a clear additional contribution at $\omega \simeq 0.45 J$ which
is found to grow and then disappears as $q_x$ is varied between $0$ and
$0.4 \pi$. This feature is readily explained by the fact that the strongest
contributions, from the $(\pi,0)$ and $(0,\pi)$ regions, have a second
characteristic energy separation related to the sum of the maximum
gap and the energy of the flat part of the spinon dispersion at $(\pi,0)$
(Fig.~4): the general denominator $\omega^2 - ( E_{k+q/2} + E_{k-q/2})^2$
in (\ref{eqdo}) remains negative at these values.

	In the $T$-dependence, the most distinctive feature is that the
contribution to $\delta \omega$ from the $u$-RVB state is no longer vanishing,
so that the anomaly at $T_{RVB}$, which is the difference between this and
the phonon self-energy in the $s$-RVB state, becomes smaller. It also appears
somewhat sharper as a function of temperature, and both of these features
agree well with the
finite-$q$ results from inelastic neutron scattering \cite{rprperh}.

	In Fig.~10 is shown the difference in the self-energy correction
between low ($T = 0.1 T_{RVB}$) and high ($T = 1.1 T_{RVB}$) temperatures
as a function of $q_x$ (a,b) and of $q$ along the (1,1) direction (c). The
peak features have a simple explanation on the basis of $E_k$: viewing the
integral to be performed as an exercise in maximizing the contact
between two spinon dispersion surfaces (one inverted for the anomalous
scattering contribution), the initial peak comes from a considerable
improvement in surface overlap around the $(\pi,0)$ points as soon as
$q$ is offset. The peak around $q / \pi \sim 0.3$ is due to overlap
between opposite sides of the low-energy part of the dispersion, which
is spanned by this wavevector, or scattering processes across the ``neck''
of the open Fermi surface. Both features are strongly dependent on the
exact shape of $E_k$, and while we suspect that the first is unlikely to
be observed in a real experiment, it is possible that the second will appear.
The contrast between Figs.~10(a) and~10(b) suggests that the latter peak
would only be discernible for the 340$cm^{-1}$ phonon mode, where the
anomalies are most pronounced due to the proximity of the mode frequency
to ``$2 \Delta$''. Comparison with the experimental results of Ref.
\cite{rprperh}, where fewer wavevector points could be sampled, shows some
correspondence to the decreasing trend seen in the $O_7$ material, although
this decrease is somewhat more rapid in the calculation. A very recent
study of the $q$-dependence of $\delta \omega$ for the 193$cm^{-1}$
$B_{2u}$ mode suggests a similar $q$-dependence for this, in that it
varies little with wavevector in the region of the zone center.

\subsection{Application to Other Modes}

	The 193$cm^{-1}$ $B_{2u}$ mode (Fig.~2) is also an out-of-phase
oscillation of $O(2)$ and $O(3)$ atoms in the plane, although in this
case the motion is in-phase between the planes of the bilayer. This mode
has recently been analyzed in $YBa_2Cu_3O_7$ by Harashina {\sl {et al.}}
\cite{rhs}, and they observe a sharp frequency shift occuring close to
the superconducting transition temperature,
with a relative magnitude $\delta \omega / \omega_0 \simeq 1\%$.
These features are in very good agreement with the results of the
single-layer model shown in Fig.~7(a) for a similar frequency $\omega_0$.
However, the authors also find that the phonon linewidth $\Gamma$ narrows
significantly below the transition, which leads to the model-independent
conclusion of a strong electron-phonon coupling. Here we have taken the
intrinsic linewidth to be a constant, independent of the spinon spectrum,
because no spinon spectral weight is expected for optic phonon frequencies
at $q = 0$, so the experimental result cannot be obtained. The correction,
Fig.~8(a), from
the imaginary part of the phonon self-energy (Fig.~3), is always positive
and non-zero for a $d$-symmetric gap choice in the approximation used.
For a quasiparticle density of states corresponding to a $d$-symmetric
gap state, it is easy to argue that in the presence of a coupling,
phonon modes of frequencies close to the maximal value of the gap will
be broadened by the quasiparticles, while those at low frequencies will
narrow, in accord with the standard picture for acoustic phonons
\cite{rssa}. However, in the present case a mechanism for the applicability
of this scenario remains to be elucidated.

	In Figs.~5-10 are shown not just results for $B_{1g}$ and $B_{2u}$
modes, but also those for modes at the same frequencies with in-phase $O(2)$
and $O(3)$ oscillations, which in the single-layer case would correspond
to $A_{1g}$ (Raman active) or $A_{2u}$ (infrared-active) symmetry. The
$A_{1g}$ mode appears at $440cm^{-1}$, above $2 \Delta$, and so if its
response was similar to $B_{1g}$ (Fig.~5(b)) would show at best a small
positive frequency shift and broadening, similar in fact to the observation
of Ref. \cite{ralt}. The $A_{2u}$ mode occurs at $307cm^{-1}$, well
positioned to show strong effects. However, one sees immediately that in
the current approximation these have negligible anomalies, a qualitative
difference from the $B$-symmetric modes which emerges from the form factors
$F_k$ (\ref{eff}) appropriate to $d$-wave singlet pairing.

	Experimentally, the situation surrounding the strongly
infrared-active $A_{2u}$ mode appears not to be clearly understood, since
neutron studies \cite{rprperh} report a ``mysteriously'' small shift $\delta
\omega / \omega_0 \simeq 0.4 \%$, whereas far infrared spectroscopy
\cite{rliv} reveals a strong shift $\delta \omega / \omega_0 \simeq 1.2 \%$.
The result from infrared reflectivity has recently been reproduced on a
single crystal of optimally-doped YBCO \cite{rtpc}, while the neutron
scattering experiment suffers in that this mode appears as an extremely
broad peak. The failure of the present model to reproduce these effects is
directly attributable to its single-plane nature: the $A_{2u}$ mode develops
a very strong dipole moment, so is accompanied by significant interplane
charge transfer, and a bilayer formulation including interband processes
in a weak-coupling scheme has already been shown \cite{rhfk} to contain
this physics. Extension of the model to a system of two coupled planes
is an area of active research, and leads to non-trivial questions about
the allowable symmetries of the superconducting gap in a bilayer which
are themselves of intrinsic interest. In the case of the $A_{1g}$ mode,
its atomic motions will also cause charge motion within the $CuO_2$ plane,
the treatment of which is beyond the basic $t$-$J$ framework.

	Finally, the effects studied will be strongly suppressed in the
$E$-symmetric phonon modes of $O(2)$ and $O(3)$, where atomic displacements
are parallel to the plane, as the modulation of $t_{ij}$ and $J_{ij}$ will
be only quadratic in the phonon coordinate ({\it {cf.}}
(\ref{elcce1},\ref{elcce2})).
In addition, we would not expect to find significant superconductive
anomalies in phonon modes involving motions of atoms in the unit cell which
are not located in the $CuO_2$ layers, with the possible exception of apical
$O(4)$, which is strongly coupled to the planar system as indicated in Ref.
\cite{rMF}. In large part these qualitative expectations are borne out by
experiment, where no other strong anomalies are observed, except indeed for
a moderate feature in
the 500$cm^{-1}$ $A_g$-symmetric mode of $O(4)$ \cite{ralt}, and a very
curious, low-frequency mode in some YBCO compounds which appears to involve
the $c$-axis motion of the $Ba$ atom \cite{rklbjb} (which we note lies also
in the plane of apical $O$).

	A concluding comment is in order on the possibilities for phonon
anomalies in other classes of high-$T_c$ ceramics. The most
extensively-studied groups of compounds are the $La_{2-x}Sr_xCuO_4$ and
related $Nd_{2-x}Ce_xCuO_4$ series with variable doping $x$, and the
parent $R_2CuO_4$ series with a selection of elements $R$; in these cases
there is now well-documented evidence for the link between superconductivity
and local structural anomalies \cite{ruea}. However, neither spectroscopic
nor neutron investigations have revealed, in the modes examined so far,
any clear signature of phonon anomalies \cite{ruapc} of the type we consider,
even in the $B$-symmetric distortion of
the single $CuO_2$ layer in the unit cell (although this occurs
above the characteristic value of ``$2 \Delta$''). While this result
is exactly that predicted by the theory, because the $CuO_2$ plane is
essentially unbuckled above $T_c$ (HTT phase), so that phonon modulation
of the $t$ and $J$ terms would be quadratic, and thus small, it is by no
means certain that the observations correspond to an intrinsic property
of the materials, rather than being an effect of impurities (section 1)
or, more fundamentally for Raman studies, the lack of true inversion
symmetry. In the LTO and LTT phases, the plane does undergo a distortion,
consisting of alternating tilts of the $CuO_6$ octahedral units in the
[1,1] and [1,0] directions, respectively, but against this background
equilibrium configuration of positive and negative $u_0$ values (section
2), the net coupling for any combination of $A$- and $B$-symmetric gap and
phonon symmetries at $q = 0$ will cancel at linear order.

	The BSCCO group of compounds also shows a buckling distortion.
While exact structural determinations are complicated by the presence of
dislocations, modulations and disorder, there have been satisfactory
characterizations of $Bi_2Sr_2CaCu_2O_8$ \cite{ryf,rtpsgs} which show
the nature of the $CuO_2$ plane buckling to be that of the LTO distortion
of the single plane in LSCO compounds. From the theory, we would expect
no observable anomalies due to the cancellation of linear terms in such
a structure (above), and this is consistent with experimental reports
\cite{rsugai}, which indicate that a frequency shift for the 336$cm^{-1}$
$B_{1g}$ mode is absent in $Bi_2Sr_2CaCu_2O_8$-related materials. Given the
sensitivity to impurities of the anomalous effects documented in YBCO
materials, these results are yet to be verified. There
remains no data on phonon anomalies also for
the $Tl$- and $Hg$-containing materials, although there has been a
detailed investigation of the local structural anomaly in the former
\cite{rtejs}.

\section{Underdoped Phase and Spin Gap}

	In the preceding sections we have detailed the predictions of a
model where the phonon anomalies are coupled to spin singlet formation,
which gives as their onset temperature not the superconducting critical
temperature $T_c$ but the $s$-RVB condensation temperature
$T_{RVB}$. This result raises the interesting possibility of probing the
spin-gap behavior found in members of the YBCO class in the low-doping
regime, by which is meant doping levels below that required for the
optimal $T_c$, by considering the frequency shifts of particular phonons.
The definition of the spin gap varies among authors and experiments, but
we take it to mean the loss of spectral weight in the spin response which
sets in at some temperature $T_0$ above $T_c$, and has been best
characterized by observations of the temperature-dependence of the NMR
relaxation rate \cite{ry}, which is maximal at this $T_0$.

	There exist already several experimental reports \cite{rliv,ralt,rham}
of anomalies in the frequency shift well above $T_c$, whose onset temperature
corresponds closely to that where the NMR rate exhibits a maximum. Recent,
highly accurate Raman scattering studies of phonon anomalies in the
stoichiometric underdoped YBCO compounds $Y_2Ba_4Cu_7O_{15}$ and
$YBa_2Cu_4O_8$ \cite{rklbjb} add considerable weight to these considerations
as they show clearly the onset of a frequency shift at some $T_0 \simeq 150K$,
followed by growth of this shift as temperature is lowered, until a saturation
below $T_c$, as shown schematically in Fig.~11.
Such features may be understood on the basis of the mean-field phase diagram
of the extended $t$-$J$ model \cite{rTKF}, reproduced in Fig.~12, in which
$T_{RVB}$ is indeed higher than $T_c$ only in the low-doping region. In this
framework, there is an onset of short-ranged singlet RVB order around $T_0$,
which is then identified with the crossover temperature $T_{RVB}$
\cite{rnlln}, and an increasing correlation length of the coherent regions
with decreasing temperature until fully coherent, long-range order is
attained at $T_c$. The saturation of the magnitude of the phonon frequency
shift below $T_c$ is a consequence of the finite-frequency nature of the
excitation (Fig.~7(b)). We speculate that studies of phonon anomalies at
higher wavevectors might show full development of the energy correction at
temperatures closer to $T_0$, but on the basis of Fig.~10 doubt that these
would be measurable.

	These results for the change in nature of the transition between low-
and high-doping regimes are consistent with the observation that
the specific heat anomalies at $T_c$ in each case are qualitatively
different \cite{rlsh,rskhs}.
The existence of two temperature scales may provide an explanation for the
contrasting low-$q$ behavior of the energy shifts in nominal $O_{6.92}$ and
$O_7$ compounds observed in Ref. \cite{rprperh}: here the authors took as
their high temperature for comparison a value of 100$K$, which if the
lower-doped compound were to possess a spin gap above $T_c$ could very well
remain in the regime with a phonon anomaly, {\it {i.e.}} below $T_0$.
We note also that there have been
some experimental reports which indicate a true transition, as opposed to a
crossover, around the higher temperature in several material classes, notably
low-doped phases of YBCO, and it is possible that this may be identified
with a lattice instability \cite{rnsmftk,rtea,rlkggsrwhkmf}.

	In conclusion to this section we comment that the present theory
contains only one temperature scale, $T_{RVB}$, and thus a detailed
quantitative explanation of the low-doping regime remains beyond its scope
(see the related note in the introduction to section 3). Within the same
framework, a more accurate account of the features reproduced in outline here
will require additional physics, perhaps in the form of an improved
treatment of the boson degrees of freedom, to restore the lower
temperature scale of the true superconducting transition.

\section{Isotope Effect}

	The isotope effect was one of the critical pieces of evidence which
pointed the way to the formulation of the BCS theory for superconductivity in
conventional metals. In the high $T_c$ materials the situation is somewhat
complex, and a comprehensive review with a wealth of experimental evidence
and a summary of theoretical models to date is given in Ref. \cite{rfier}.
In brief, the qualitative trend in each material class appears to be a
situation where the oxygen isotope effect, parameterized by $\alpha =
- d \ln T_c / d \ln M \simeq - \mbox{$\frac{\Delta T_c}{T_c} \frac{M}{\Delta
M}$}$, is close to vanishing at the optimal doping level,
but rises smoothly to values of at least $\alpha = 0.5$ on sufficient under-
and over-doping. The question of sample purity and randomness away from the
stoichiometric compounds (section 1) may be pertinent here once again,
but this form has been observed in almost all types of system.
In general, electronic models such as the current one do
not address the question of the isotope effect at all, and in this work we
will have the modest aim only of elucidating the influence of the spin-phonon
coupling at the near-optimal doping level.

	The thermodynamic properties of the coupled system may be studied
self-consistently by including phonon-spinon terms in the free energy $F$
(Fig.~13(a)). In a Ginzburg-Landau expansion, the free energy difference
between a normal state and a superconducting one with a single order parameter
$\Delta$ has the form $F_S - F_N = A(T) \Delta^2 + B(T) \Delta^4 + \dots$ It
is tempting to postulate that the $\Delta^2$ contribution of the additional
parts acts to enhance the stability of the spin singlet state, providing a
natural linkage of the spin gap to the lattice structure which might
account for the variety of characteristic temperatures discussed in the
previous section. However, the relative magnitude of the effective
phonon-spinon vertex may be estimated as shown at the end of section 2,
and because the contributions due to fluctuations in $t$ and $J$ require
two such vertices, their effect can be expected to be of order $10^{-2}$.
Here we estimate this effect by evaluating the $\Delta^2$ contributions
from the lowest-order phonon-spinon coupling term at $T_{RVB}$, using
the spinon propagators for the normal, or $u$-RVB, state.

\begin{table}[p]
\centering

\begin{tabular}{|c||c|c|c|c|}
\hline & (i) & (ii) & (iii) & total \\
\hline
\hline  $A_{1g}$ & 1.545 $\times$ 10$^{-3}$ & 3.453 $\times$ 10$^{-4}$
& -3.588 $\times$ 10$^{-4}$ & 1.532 $\times$ 10$^{-3}$ \\
\hline  $B_{1g}$ & 1.319 $\times$ 10$^{-3}$ & -1.775 $\times$ 10$^{-4}$
& 3.210 $\times$ 10$^{-3}$ & 4.351 $\times$ 10$^{-3}$ \\
\hline  $A_{2u}$ & 1.812 $\times$ 10$^{-3}$ & 3.755 $\times$ 10$^{-4}$
& -4.223 $\times$ 10$^{-4}$ & 1.765 $\times$ 10$^{-3}$ \\
\hline  $B_{2u}$ & 1.718 $\times$ 10$^{-3}$ & -2.558 $\times$ 10$^{-3}$
& 4.547 $\times$ 10$^{-3}$ & 3.708 $\times$ 10$^{-3}$ \\
\hline
\end{tabular}


\caption{Lowest-order spinon-phonon coupling contributions to the $s$-RVB
transition temperature. The magnitude is relative to $T_{RVB} = 0.069 J$,
and the sign is chosen so that a positive contribution corresponds to an
enhancement of $T_{RVB}$. Figures are given at $T_{RVB}$ for the 4 modes
involving $c$-axis motion of planar $O(2)$ and $O(3)$ atoms (Fig.~2), and
for the 3 types of process shown in Fig.~13(b). }
\end{table}

	Omitting details of the lengthy but straightforward calculation,
the free energy contributions from phonon coupling are shown in Fig.~13(b)
for the terms (i), the $s$-RVB vertex part, (ii), a pair of diagrams
corresponding to a spinon self-energy correction by the phonon, and (iii)
phonon vertex correction of the $u$-RVB vertex. Near the transition,
$F_S - F_N$ may be linearized in temperature, so that the coefficient
of the $\Delta^2$ term becomes $A(T) = a ( T - T_{RVB})$. The phonon
contributions are approximately $T$-independent, so have the schematic
form $ - b \Delta^2$, and can be written as a constant shift $\delta
T_{RVB} = b / a$; in Table~1 are given the values of the relative shifts
$\delta T_{RVB} / T_{RVB}$ arising from each of the diagram types in
Fig.~13(b), and for the four phonon symmetries and frequencies (Fig.~2).
One observes from the figures that the anomalous vertex part remains
approximately constant for all of the modes, while each of the other two
parts is small for the $A$-symmetric modes, but has large contributions
from the $B$-symmetric ones, as would be expected on the basis of the
phonon anomalies computed in section 3. The figures can be shown to
be approximately constant by performing the same calculation at different
temperatures, using the slightly different self-consistent values of the
$u$-RVB order parameter $\bar{\chi}$ and chemical potential $\mu$ at each.
The total free energy shift for these four modes of a bilayer is compared with
twice the free energy of the single-layer system \cite{rTKF}. The net
contribution corresponds to a 1.1\% enhancement of
$T_{RVB}$, a magnitude exactly in line with the qualitative expectation,
which to the same degree of accuracy is a 1.2$K$ enhancement. The isotope
shift for $^{16}O \rightarrow ^{18}O$ on this quantity is approximately
-6\%, leading to the prediction of a total shift due to in-plane oxygen
of -0.07$K$. The equivalent $\alpha$ parameter is 0.005. We would expect the
quantitative agreement to be improved when interplane hopping processes
are taken into account (section 3), in order to describe correctly the
contributions from the $A$-symmetric modes (Table~1).

	This result may be compared with a recent site-selective substitution
study of YBCO near
optimal doping, performed by Zech {\sl {et al.}} \cite{rzea}.
These authors were able to achieve high exchange rates of $^{16}O$ by
$^{18}O$ during crystal preparation, while discerning with good resolution
the location of the exchange, whether it was on planar $O(2)$ and $O(3)$
sites, or apical $O(4)$ and chain $O(1)$ sites. They measured a total $O$
isotope shift (complete exchange) of -0.25$K$, of which at least 80\%,
or -0.20$K$, was found to be due to modes of in-plane oxygen. Qualitatively,
this conclusion contains important agreement with the current picture, that
it is the planar $O$ modes which are most strongly coupled to the
superconducting interaction. At the quantitative level, the theoretical
result from the $c$-axis modes appears to be too small by a factor greater
than 2, which, while constituting an acceptable level of agreement given the
nature of the model, is not in as good accord as the phonon anomaly results
within the same approximation.

	We have shown that a realistic spin-phonon coupling within an
electronic model for superconductivity can show the small isotope effect
consistent with experimental observation in the high-$T_c$ superconducting
ceramics at optimal doping. In the present theory we would not expect to see
significant alterations in the magnitude of this shift on changes in doping,
which will act to move the chemical potential and cause minor changes in the
spinon dispersion (Fig.~4) away from the optimally-doped level. From the
phase diagram discussed in section 4, one observes that $T_c$ falls with
increasing doping in the overdoped region, where it is given by $T_{RVB}$,
and with decreasing doping in the underdoped region, where it is given by
$T_B$ (Fig.~12). With $\Delta T_c$ ($\Delta T_{RVB}$) approximately constant,
the threefold reductions in $T_c$ achieved experimentally by overdoping in
some systems \cite{rfier} would give a commensurate increase in $\alpha$;
in the underdoped regime, the changes in $T_B$ with doping have not been
estimated.

	From these considerations, it seems unlikely that the $t$-$J$
framework alone contains the origin of the spectacular changes in isotope
effect with doping reported in experiment, or that a purely electronic
picture could account for the apparent changeover to large values ($\simeq
0.5$) of $\alpha$ which these data present. However, the situation is
complicated due to questions of $O$ disorder, and even equilibrium position,
on doping and isotope substitution, the actual fraction of $O$ in the plane
sites, and the sensitivity of charge fluctuations to these uncertainties.
At minimum, we may suggest that the results show once again the spin-phonon
interaction to be a detectable
effect which can be used to probe the physics of the high-$T_c$ materials,
but to be only a small perturbation on the dominant processes. In the case
of the isotope effect, this perturbation is seen clearly at the optimal
doping level where the dominant contribution vanishes, but a full
explanation of the latter will require additional or separate physics. Ideas
which have been put forward \cite{rfier} include
van Hove singularities, interlayer pair tunneling, bipolaron formation,
strong Coulomb correlations and isotope-dependent hole concentration,
but many models based on these contain a conventional, often phenomenological,
electron-phonon interaction, and suffer from other weaknesses which this
entails.

\section{Conclusion and Discussion}

	We have proposed a theory of spin-phonon coupling, based on the
mean-field approximation to the extended $t$-$J$ model of a single $CuO_2$
plane, which provides good agreement with a variety of experimental
observations on high-$T_c$ cuprates of the YBCO class. A key element is
the buckling of the $CuO_2$ plane present in these materials, which causes
the coupling to be linear in $O$ displacement along the $c$-axis.
The model contains the following features:

\smallskip
\noindent
i) it gives, without recourse to parameter-fitting, a semi-quantitative
account of the frequency and linewidth anomalies observed in the important
and well-characterized $B_{1g}$- and $B_{2u}$-symmetric
phonon modes of planar oxygen atoms in the stoichiometric $O_7$ compound.

\smallskip
\noindent
ii) the symmetry of the superconducting gap consistent with experiment
is found to be predominantly $d_{x^2 - y^2}$.

\smallskip
\noindent
iii) the conclusions are supported by the results obtained for different
phonon mode types, and from observations in other materials.

\smallskip
\noindent
iv) those features of the phonon anomalies which are not given by the
present treatment are expected to be readily described within the same
framework by consideration of a bilayer structural unit, and by a more
sophisticated treatment of the intrinsic phonon linewidth.

\smallskip
\noindent
v) at lower doping levels, away from that for optimal $T_c$, there is good
qualitative agreement, which provides additional strong evidence, consistent
with the results from NMR and inelastic neutron scattering, for the origin
of the spin gap,

\smallskip
\noindent
vi) there is a satisfactory degree of correspondence with the measured
isotope shift near optimal doping, and the phonon modes giving the most
important contributions to this are identified.

\smallskip
\noindent
The single-layer $t$-$J$ model for the strongly-correlated $CuO_2$ system
has thus been shown to contain an additional class of physical phenomena,
namely those related to structural anomalies. The theory is found to be
well-suited to illustrating a variety of interesting properties observed
in experiment, and to constitute a basis for further development.

	Based on the results for alteration of phonon dynamics by spin
excitations, it would seem logical also to seek the alteration of spin
dynamics by their interaction with the phonon degrees of freedom. In fact this
study was motivated initially by the failure of the extended $t$-$J$ model to
reproduce at the mean-field level some features of the YBCO spin excitation
spectrum, notably the 41$meV$ peak in the $O_7$ compound
\cite{rrrvbbbhl,rmyama}. The close
coincidence of this resonance energy with the frequency of the $B_{1g}$
mode, whose superconducting anomalies are the strongest, and the
$T$-dependence of the peak intensity, which scales with $\Delta^2 \propto
\rho_s$, the superfluid density, make the present spin-phonon coupling
theory appear well-suited to explain these features. However, as in sections
3 and 5, the phonon contributions to the spin susceptibility through the
spin-phonon coupling vertices of section 2 are of order
a few percent, while the 41$meV$ peak dominates the entire spin response
\cite{rmdamhhasl}. One may consider instead direct coupling of the phonon to
spin-flip processes via the spin-orbit (Dzyaloshinskii-Moriya) interaction
in the buckled $CuO_2$ plane \cite{rnb}, but this will be 2 orders of
magnitude smaller still. We note also that the exact position of the
resonance depends on the doping $\delta$, with the energy declining on moving
to lower doping \cite{rrs}, while the frequency of the $B_{1g}$ mode appears
to remain approximately constant. In addition, there is no trace of an
analogous feature in the spin response associated with the $B_{2u}$ phonon
mode, which also shows a considerable anomaly.

	While we have explored in some detail the consequences of a
spin-phonon coupling in this class of model, many questions remain. One
of the major shortcomings of the theory is its failure to deal systematically
with a physical range of hole-doping levels, as explained in sections 3 and
5, but despite this it does provide a consistent description of
the spin gap, as discussed in section 4. Another issue which
remains a drawback of the slave-boson formulation is the treatment of the
holon degrees of freedom: in the model they have an intrinsic tendency towards
Bose condensation, and thus contribute little to the dynamical phenomena, even
within a gauge-field scheme where the appropriate spin- and hole-backflow
requirements are enforced, leading to the Ioffe-Larkin compostion rules for
transport properties \cite{rIL,rnlln}. Some studies indicate that the holes
also possess a Fermi surface \cite{rPutikka,rcl}, which suggests that the
appropriate physical description will require hard-core bosons or secondary
statistical transmutation. In the present context, this would lead to
the introduction of a meaningful second temperature scale in the low-doping
regime (section 4), and the possibility of a realistic account not only of
the development of the phonon anomalies, but also of the features in the
spin response detailed by NMR and inelastic neutron scattering experiments.

	To elaborate on the observable consequences of a gauge-theory
approach to describing fluctuations around a mean-field solution, we have
considered the possibility of gauge-field screening \cite{rnlln} of the
phonon anomalies. This is discussed briefly in the appendix for the massive,
longitudinal gauge modes, which can in principle have a Fano-type
coupling to phonons. We show that the corrections due to gauge fluctuations
are small or vanish, depending on the phonon mode symmetry.
However, the effects of the massless, transverse gauge modes in providing
quasiparticle self-energy and vertex corrections have been studied only
briefly in the context of their influence on the dynamical properties of the
spin degrees of freedom \cite{rFK}, and remain an open question.

	In conclusion, we propose that the model discussed represents
a useful step in the formulation of a coherent theory of spin excitations,
transport properties and lattice degrees of freedom. The agreement between
theory and experiment achieved at the current level is encouraging in
efforts to construct a unified picture of the anomalous metallic state,
whose characterization is vital to a full understanding of high-temperature
superconductivity.

\section*{Acknowledgements}

  We are grateful to M. Arai, T. Arima, Y. Endoh, H. Harashina, H. Kino,
N. Nagaosa, H. Oyanagi, G. Sawatzky, S. Tajima, Y. Tokura, M. Udagawa,
H. Yasuoka and especially H. Eskes and
M. Sato for helpful discussions. H.F. would particularly like to thank
B. Batlogg for stimulating conversations during the Workshop on
Cuprate and Heavy-Fermion Superconductors in Cologne (Sept. 1993).
This work was supported financially by the Monbusho International
Scientific Research Program: Joint Research ``Theoretical Studies on
Strongly-Correlated Electron Systems'' (05044037) and the Grant-in-Aid for
Scientific Research in Priority Areas ``Science of High-$T_c$
Superconductivity'' (04240103) of the Ministry of Education, Science and
Culture, Japan. B.N. wishes to acknowledge the support of the Japan
Society for the Promotion of Science.

\section*{Appendix}
\setcounter{equation}{0}
\renewcommand{\theequation}{A\arabic{equation}}

	Fluctuations about the mean-field solution of the order parameters
in a model of the type we consider may be taken into account by the
introduction of a gauge field \cite{rnlln},
whose transverse components, corresponding to phase fluctuations, turn out
to be massless, and whose longitudinal component, corresponding to density
fluctuations, is massive. Without going into the full details of the
gauge-field formulation, we may consider the effect of possible gauge-field
screening in the spin-phonon coupling problem as follows.

	The bare gauge field is merely a restatement of boson-fermion
coupling within the slave boson decomposition, and has no dynamics, but
the boson and fermion polarization terms, occurring in sequences of any
length, act to generate effective dynamical properties \cite{rnlln}.
Restricting the discussion to Fano-type processes, the
full gauge-field correction by polarization terms to the lowest-order
spinon-phonon diagram shown in Fig.~3 is then simply the diagram in
Fig.~A1(a), where the dotted line is the full dynamical gauge field
propagator $- 1 / (\Pi_B + \Pi_F)$. Because the phonon vertex, from $t$ and
$J$ terms, contains only spin-spin and density-density components, it may
couple only to the longitudinal part of the gauge field, whose propagator is
simply that of a mass $M_0 \sim J$, while the magnitude of the gauge-field
vertex is approximately $J$; these considerations may be made more
quantitative, but the important feature of the result is unaffected. As
only the spinon coupling to the phonon is considered, each end ``bubble''
in the full gauge-field-mediated sequence, with one $k$-dependent phonon
and one constant gauge vertex, may be denoted by $\Pi_{sp}^{\prime}$, in
which case the total corrected expression at the order of Fig.~3 is
$\Pi_{sp} - \Pi_{sp}^{\prime 2}$.

	Brief examination of the diagram (Fig.~A1(b)) for $\Pi_{sp}^{\prime}$
shows that at $q = 0$ the expression must be $O(\Delta^2)$: with a $u$-RVB
vertex the anomalous propagators are required, while with an $s$-RVB vertex,
of $O(\Delta)$, a single anomalous propagator appears, so the correction
will unavoidably be $O(\Delta^4)$, and unimportant. One may write down an
expression analogous to (\ref{edo}) for {\sl {Re}}$\, \Pi_{sp}^{\prime}$,
and find that the form factor is
\begin{eqnarray}
F_{k}^{\prime} & = & 2 \Delta^2 \eta_k \left( \gamma_k \xi_{k} +
\mbox{$\frac{3 J}{4}$} {\bar {\chi}} \eta_{k}^2 \right) \;\;\;\;\;\;\;\;\;\;
\;\;\; B_{1g}, \, B_{2u} \; {\rm {modes}} \label{effgfc} \nonumber \\
F_{k}^{\prime} & = & 2 \Delta^2 \eta_{k} \left( \eta_k \xi_{k} +
\mbox{$\frac{3 J}{4}$} {\bar {\chi}} \gamma_k \eta_k \right)
\;\;\;\;\;\;\;\;\;\; A_{1g}, \, A_{2u} \; {\rm {modes}} .
\end{eqnarray}
This is odd in $\eta_k$ for the $B$-symmetric modes (in fact for both $s$-
and $d$-symmetry of the singlet order parameter), and so vanishes upon
$k$-integration, while for $A$-symmetric modes it remains of the order of
$\Pi_{sp}$, and so becomes irrelevant on taking the square.
Thus the consequence of considering fluctuations through a gauge-field
coupling, which will involve only massive longitudinal modes with no
singularities in momentum space, is that their correction is small or
vanishing.

\newpage

\section*{Figure Captions}

\noindent
Fig. 1: Schematic representation of the $CuO_2$ layer. a) Planar structure.
b)``Buckling'' deformation of the equilibrium positions of $O(2)$ and $O(3)$
atoms out of the plane of the $Cu$ atoms, appropriate for YBCO. $a$, $b$ and
$c$ represent the crystal axes (assuming tetragonal symmetry).

\bigskip
\noindent
Fig. 2: Schematic representations of phonon modes involving $c$-axis
oscillations of in-plane oxygen atoms in the YBCO bilayer structural
unit: a) $A_{1g}$, b) $B_{1g}$, c) $A_{2u}$ and d) $B_{2u}$.

\bigskip
\noindent
Fig. 3: Diagrammatic form of the lowest-order contribution
to the phonon self-energy $\Pi_{sp}$ due to coupling to spinons. The
thick line denotes the spinon propagator in Nambu representation, and the
dot the corresponding 2$\times$2 spinon-phonon vertex.

\bigskip
\noindent
Fig. 4: Energy dispersion $E_k = - \sqrt{ \xi_{k}^2 + \Delta_{k}^2 }$ for
a YBCO-like system, where the transfer integrals in the extended
$t$-$J$ model are $t = 4J$, $t^{\prime} = - \frac{1}{6} t$ and $t^{\prime
\prime} = \frac{1}{5} t$, corresponding respectively to nearest-, second-
and third-neighbor hopping processes.

\bigskip
\noindent
Fig. 5: Phonon frequency shift $\delta \omega$ for $B_{1g}$ ($\circ$)
and $A_{2u}$ ($\times$) modes at $q = 0$ as a function of frequency at
a) $T = 0.2 T_{RVB}$ and b) $T = 0.8 T_{RVB}$. The arrows indicate the
frequencies whose $T$-dependence is illustrated in Fig.~7.

\bigskip
\noindent
Fig. 6: Phonon linewidth broadening $\delta \Gamma$ for $B_{1g}$ ($\circ$)
and $A_{2u}$ ($\times$) modes at $q = 0$ as a function of frequency at
a) $T = 0.2 T_{RVB}$ and b) $T = 0.8 T_{RVB}$. The arrows indicate the
frequencies whose $T$-dependence is illustrated in Fig.~8.

\bigskip
\noindent
Fig. 7: Phonon frequency shift $\delta \omega$ for $B_{1g}$ ($\circ$)
and $A_{2u}$ ($\times$) modes at $q = 0$ as a function of $T$
for mode frequencies a) $\omega_0 = 0.15 J$ and b) $\omega_0 = 0.25 J$.

\bigskip
\noindent
Fig. 8: Phonon linewidth broadening $\delta \Gamma$ for $B_{1g}$ ($\circ$)
and $A_{2u}$ ($\times$) modes at $q = 0$ as a function of $T$,
for mode frequencies a) $\omega_0 = 0.15 J$ and b) $\omega_0 = 0.25 J$.

\bigskip
\noindent
Fig. 9: Phonon frequency shifts $\delta \omega$ for $B_{1g}$ ($\circ$)
and $A_{2u}$ ($\times$) modes at wavevector at ${\bf q} = (0.2,0) \pi$;
a) as a function of frequency at $T = 0.2 T_{RVB}$ and b) as a function
of $T$ for mode frequency $\omega_0 = 0.25 J$.

\bigskip
\noindent
Fig. 10:  Phonon frequency shift $\delta \omega$ for $B_{1g}$ ($\circ$)
and $A_{2u}$ ($\times$) modes of wavevectors a) $q_x$ in the $(1,0)$
direction, for a phonon of bare frequency $\omega_0 = 0.15 J$, b) $q_x$
in the $(1,0)$ direction, for a phonon of bare frequency $\omega_0
= 0.25 J$, and c) $q / \sqrt{2}$ in the $(1,1)$ direction, for a phonon of
bare frequency $\omega_0 = 0.25 J$.

\bigskip
\noindent
Fig. 11: Schematic representation of the frequency shift of the 340$cm^{-1}$
$B_{1g}$ phonon mode as a function of temperature in the underdoped compound
$YBa_2Cu_4O_8$, from the data of Ref. \cite{rklbjb}. $T_c = 82K$ is the
temperature of the superconducting transition, and $T_{sg} \simeq 150K$
denotes the onset of the anomalous frequency shift.

\bigskip
\noindent
Fig. 12: Mean-field phase diagram for the extended $t$-$J$ model in the
slave-boson formulation, showing the existence of a spin gap at doping
levels below that for the optimal $T_c$ (after Ref. \cite{rTKFs}).

\bigskip
\noindent
Fig. 13: Diagrammatic representation of a) the free energy contribution due
to the lowest-order spinon-phonon couping term, and b) those parts of order
$\Delta^2$, to be evaluated in the normal state at $T_{RVB}$.

\bigskip
\noindent
Fig. A1: Gauge-field correction to the lowest-order spinon polarization.
a) Representation of all orders of correction due to gauge field. b)
End bubbles in diagrammatic sequence, showing $O ( \Delta^2 )$ nature.

\end{document}